\documentclass[preprint,amsmath,amssymb]{revtex4}
\usepackage{upgreek}
\usepackage{graphicx}
\usepackage[dvipsnames]{xcolor}
\usepackage{subcaption}
\usepackage{tabularx}
\usepackage{multirow}
\usepackage{array}
\usepackage{makecell}
\usepackage{dcolumn}
\usepackage{adjustbox}
\usepackage{hyperref}
\usepackage{bm}

\begin{document}
\title
{Conductors and superconductors in stationary spacetimes: generalized Schiff-Barnhill and Meissner effects}
\author{${\rm F.\;Masghatian}^{\;(a)}$\footnote {Electronic
address:~fa.masghatian@ut.ac.ir}, ${\rm M.\;Nouri}$-${\rm Zonoz}^{\;(a)}$\footnote{Electronic address:~nouri@ut.ac.ir \; (Corresponding author)} and ${\rm H. \;Sadegh}^{\;(b,c)}$ \footnote{Electronic
address:~Hamed.Sadegh@tum.de}}
\affiliation{(a): Department of Physics, University of Tehran, North Karegar Ave., Tehran 14395-547, Iran.\\
(b): Technical University of Munich, School of Natural Sciences, Physics Department, James-Franck-Straße 1, 85748 Garching, Germany.\\
(c): Ludwig-Maximilians Universität, Arnold Sommerfeld Center for Theoretical Physics, Theresienstr. 37, 80333 Munich, Germany.}
\begin{abstract}
Previous studies on conductors  and  superconductors in curved spacetimes, and the gravitationally-induced electric and magnetic fields in them, are  mostly restricted to the limit of weak fields and low velocities. Also the observer-dependence of these phenomena are not discussed in the literature. In this article, following London's phenomenological approach to superconductivity, we consider the electrodynamics of superconductors in stationary spacetimes without employing any weak field or low velocity limit. To this end, first we define the 3-force acting on charged particles in gravitational and electromagnetic fields, and derive its explicit form  in a general spacetime decomposition formalism. Then, to highlight the role of observers, we apply the force equation to the free charges in superconductors in  stationary spacetimes in the two well known threading and slicing  decomposition formalisms. Applying the stationary condition in the superconducting state, we arrive at the generalized Schiff-Barnhill and Meissner effects, which compared to the low velocity and weak field limit include extra terms originating from the recently introduced gravitationally-induced constitutive equations \cite{MEN}. To show the consistency of our approach, we also derive the generalized Meissner effect from the second London equation in stationary spacetimes.
\end{abstract}
\maketitle
\section{Introduction}
The study of electromagnetic fields inside conductors, and superconductors in the presence of gravitational fields started with the pioneering works of Schiff and Barnhill \cite{Schiff}, and DeWitt \cite{DeWitt} respectively. Schiff and Barnhill have shown that if the conductor is in a gravitational field or is accelerated by an acceleration $ - \bf g$ (employing the equivalence principle), it is not the electric field that vanishes inside the conductor in electrostatic equilibrium, but the following combination of fields,
\begin{equation}\label{F1}
{\bf E} + (m/q){\bf g} = 0,
\end{equation}
in which $q$, and $m$ are the charge and mass of the mobile charges. In other words there will be a gravitationally induced electric field inside  a conductor in a gravitational field.\\
DeWitt pointed out that in the presence of stationary gravitational fields, producing a gravitational drag, there should be a similar effect corresponding to induced magnetic field inside superconductors.
Starting from the Lagrangian of a charged particle with mass $m$ and charge $q$, in an electromagnetic field $A_\mu$
\footnote{The Greek indices run from $0\; {\rm to}\; 3$, while the Roman ones run from 1 to 3. Also we use units in which  $c = G =1$.}, in a curved background with metric $g_{\mu\nu}$, namely
\begin{equation}
{L}=-m~(g_{\mu \nu} \dot{x}^{\mu}\dot{x}^{\nu})^{\frac{1}{2}} - {q}A_{\mu}\dot{x}^{\mu}, \label{Lag}
\end{equation}
he arrives at the following Hamiltonian for the  particle in the low velocity and weak field limit (i.e $v \ll 1$ and,  $g_{\mu\nu} = \eta_{\mu\nu} + h_{\mu\nu}, h_{\mu\nu} \ll 1$ ),
\footnote{In DeWitt's derivation of the full Hamiltonian there is an error which was corrected in later studies (\cite{Cognola}). Apparently this does not affect his weak field and low velocity Hamiltonian, though it might produce some confusion in distinguishing between canonical and mechanical momenta.},
\begin{equation}\label{Ham}
{H}=\frac{1}{2m}({\bf P} + q{\bf A} - m{\bf h_0} )^2 -(q A_0+\frac{1}{2}m h_{00}),
\end{equation}
in which ${\bf P}$ is the canonical momentum, and $h_{00}$, and ${\bf h_0}=(h_{01}, h_{02}, h_{03})$ are the small corrections to the Minkowski metric representing a weak stationary gravitational field.
Generalizing the above Hamiltonian to the case of an ensemble of free particles (electrons) inside a superconductor, and assuming that the machinery of BCS theory can be applied to this Hamiltonian, he arrives at the following combination of fields vanishing inside a superconductor as a {\it modified Meissner effect},
\begin{equation}\label{F2}
-q \nabla \times  {\bf A} + m \nabla \times  {\bf h}_0 = 0 \;\;\;\; or \;\;\;\;\; q {\bf B} + m{\bf B_g} = 0
\end{equation}
where in the second form of the equation we have written the second term in terms of the definition of the so called {\it gravitomagnetic} field of the stationary spacetime, which in the weak field limit is given by  ${\bf B}_g = - \nabla \times  {\bf h}_0 $ (refer to section III and IV). By the above equation, DeWitt shows it is the flux of this quantity which is quantized through a superconducting ring. Indeed by writing the approximate Hamiltonian \eqref{Ham}, which is quadratic in mechanical momentum ${\bf p} \equiv {\bf P} - q{\bf A} + m{\bf h_0}$, he simply reads the generalized vector potential $- q{\bf A} + m{\bf h_0}$ whose curl gives the vanishing field inside a superconductor in a weak stationary gravitational field.
Several authors \cite{Pap, Hira, Ross, Annan, Cognola1, Li} following DeWitt, studied the same effect in different settings, and through different approaches either in the weak field, and low velocity limit, or without employing the  decomposition in terms of the gravitoelectromagnetic, and electromagnetic fields in different decmpositions formalisms.\\
Here instead of using the Hamiltonian approach, as DeWitt did, we follow London's phenomenological approach,  based on the equation of motion of a superconducting electronic fluid, to arrive at the general form of the above two effects, i.e,  without applying weak field, or low velocity limit. To do so we start from the formulation of force equation acting on a charged free particle in gravitational, and electromagnetic fields. The force equation is introduced in the context of spacetime decomposition formalisms, and their corresponding observers. To highlight the role of the observers, we  employ the well known $1+3$ ({\it threading}), and $3+1$ ({\it slicing}) spacetime decomposition formalisms corresponding to the so called fundamental and fiducial observers respectively. Indeed to the best of our knowledge, this is the first study in which the role of observers is taken into account in such phenomena.
It is shown that in our case the modifications in the Schiff-Barnhill, and Meissner effects originate from the  gravitationally-induced {\it constitutive equations} \cite{MEN}, in the corresponding decomposition formalism.\\
The outline of the paper is as follows. In the next section  we define  the 3-force acting on a charged particle in electromagnetic, and gravitational fields, and  give its explicit form in a general  decomposition formalism. In sections III and IV we calculate this  3-force in the context of threading, and slicing decomposition formalisms respectively. In section V we employ London's approach, and show how one can arrive at the generalized Schiff-Barnhill, and Meissner effects without applying weak field or low velocity limit. We show the consistency of our approach by calculating the generalized Meissner effect from second London equation in the presence of a stationary gravitational field. In the concluding section, we  discuss and compare our results in the two decomposition formalisms.
\section{Force acting on a charged particle in gravitational and electromagnetic fields}
The concept of force was central in the formulation of the laws of motion by Newton, the concept he also employed to explain the gravity and free fall. The same notion was later used in the study of electromagnetic phenomena, and specifically in the introduction of the  electric force in direct analogy with the gravitational force. On the other hand introducing the notion of 3-force in the context of general relativity, may look like a downgrade from the 4D covariant formulation of GR. But from an observational point of view that is a necessity since  our observations include 3D objects, such as spatial distances, and 3-velocities. \\
Obviously in the context of GR, any definition of 3-force should take into account the equation of motion of a particle in gravitational and electromagnetic fields. Starting from the Lagrangian \eqref{Lag} written in terms of the particle's 4-velocity $u^\mu = \frac{dx^\mu}{ds}$, namely
\begin{equation}
{L}=-m~(g_{\mu \nu} u^{\mu} u^{\nu})^{\frac{1}{2}} - {q}A_{\mu}u^{\mu}, \label{Lagu}
\end{equation}
and using the  Euler-Lagrange equation, we end up with the following equation of motion of a charged particle in gravitational, and electromagnetic fields \footnote{To distinguish between the 3D and 4D absolute derivatives, they are denoted by ${}^3D$, and ${}^4D$ respectively},
\begin{equation}
\frac{{}^4D u^{\mu}}{ds} \equiv \frac{du^\mu}{ds} + \Gamma^\mu_{\beta\nu} u^\beta u^\nu =\frac{q}{m} F^{\mu\beta}u_{\beta}\label{1},
\end{equation}
where $F_{\mu \nu}$ is the electromagnetic field tensor.
Obviously to have a consistent  definition of  a 3-force acting on a charged particle in an electromagnetic field in a curved background, one should take into account the above equation. On the other hand such a 3D definition requires a decomposition formalism so that its corresponding observer could define spatial distances, and time intervals, and accordingly, a 3-velocity and a 3-momentum.\\
We  define the total 3-force (both gravitational and electromagnetic), acting on a charged particle in gravitational, and electromagnetic fields, as the 3D absolute derivative of the particle's 3-momentum with respect to the {\it proper time} defined in the decomposition formalism, i.e,
\begin{equation}\label{FORCE}
{\boldsymbol f }=\frac{{}^3D{\bf p}}{d\tau}.
\end{equation}
In component form, and for $p^i = m \upgamma v^i $ with $m$  the particle's rest mass, it is given by
\begin{equation}\label{FCF}
{f}^{i} =\frac{{}^3 D p^i}{d\tau}= m\frac{{}^3 D (\upgamma v^i)}{d\tau}
\end{equation}
where $v^i$, and $\upgamma
= \frac{1}{\sqrt{1-{v}^2}}$, are respectively, the components of the  particle's 3-velocity, and the Lorentz factor {\it defined} in the corresponding decomposition formalism. \\
In a general  decomposition formalism the spacetime metric is given by
\begin{align}\label{nf1}
ds^2 = g_{\mu\nu} dx^\mu dx^\nu =  {d\tau}^2- dl^2 = c^2 {d\tau}^2  -\gamma_{ij}dX^{i}dX^{j}
\end{align}
in which $\gamma_{ij}$ is the metric of the 3D space, and $d\tau$, and $dl$ are the proper time, and the spatial distance defined in the corresponding decomposition formalism \footnote{It is noted that $\upgamma$-factor should not be confused with the determinant of the spatial metric $\gamma_{ij}$ which will be denoted by $\gamma$. In the next two sections we will use indices $t$ and $s$ for this factor to distinguish it between the threading and slicing decomposition formalisms.}. These spacetime decomposition formalisms are  in turn defined as local projection of geometric quantities along different observer 4-velocities \cite{MEN}. Based on these definitions, one can assign a 3-velocity to a particle using the above line element by writing it as
\begin{align}\label{nf2}
ds^2 = {d\tau}^2 (1 -\gamma_{ij}\frac{dX^{i}}{ d\tau}\frac{dX^{j}}{ d\tau}) =  {d\tau}^2 (1 -\gamma_{ij}{v^{i}}{v^{j}})
\end{align}
where the 3-velocity is defined by
\begin{equation}\label{nf3}
{v}^i = \frac{dX^i} {d\tau}.
\end{equation}
Also it is noted that
\begin{equation}\label{nf4}
{ds}  = \sqrt{1-{{v}^2}} ~d\tau \equiv \frac{1}{\upgamma} ~d\tau.
\end{equation}
Employing the above definitions, the force equation \eqref{FCF} is given by
\begin{equation}
{f}^{i} = (\frac{m} {\upgamma})\; \frac{{}^3 D (\upgamma v^i)}{ds} = \frac{m} {\upgamma} \left(\frac{d (\upgamma v^{i})}{ds} + \upgamma \lambda^i_{\;j k} v^{j}u^{k}\right), \label{force1}
\end{equation}
in which $\lambda^i_{\;j k}$ are the components of the 3D Christoffel symbol made out of the 3D space metric $\gamma_{ij}$ as,
\begin{equation}\label{Lamb}
\lambda^{i}_{jk}=\frac{1}{2}\gamma^{im}\left\lbrace  \partial_{j}\gamma_{mk}+\partial_{k}\gamma_{jm}-\partial_{m}\gamma_{jk}\right\rbrace.
\end{equation}
Depending on the relation between the spatial components of the 4-velocity, and the defined 3-velocity sometimes it is easier to work directly with the lowered-index force equation, which by noting that $p_i= \gamma_{ij}p^j = m\upgamma v_i$, is given by
\begin{equation}
{f}_{i} = (\frac{m} {\upgamma})\; \frac{{}^3 D (\upgamma v_i)}{ds} = \frac{m} {\upgamma} \left(\frac{d (\upgamma v_{i})}{ds} - \upgamma \lambda_{i,\;j k} v^{j}u^{k}\right) = \frac{m} {\upgamma} \left(\frac{d (\upgamma v_{i})}{ds} - \frac{1}{2}\upgamma \gamma_{j k, i} v^{j}u^{k}\right), \label{force2}
\end{equation}
where $\lambda_{i,\;j k} \equiv \gamma_{in}\lambda^n_{j k} $. On the other hand taking the spatial components of \eqref{1}, in terms of the components of the particle's 4-velocity $u^\mu = \frac{dx^\mu}{ds}$, we have,
\begin{equation}\label{eom}
\frac{{}^4D u^{i}}{ds} =\frac{du^i}{ds} + \Gamma^i_{\mu\nu} u^\mu u^\nu =\frac{q}{m} F^{i\mu}u_{\mu},
\end{equation}
the lowered index form of which is given by
\begin{equation}\label{eoml}
\frac{{}^4D u_{i}}{ds} =\frac{du_i}{ds} - \Gamma_{\mu, i \nu} u^\mu u^\nu = \frac{du_i}{ds} - \frac{1}{2}g_{\mu, i \nu} u^\mu u^\nu =\frac{q}{m} F_{i\mu}u^{\mu}.
\end{equation}
By finding spatial components of the 4-velocity $u^i$ ($u_i$), in terms of the 3-velocity $v^i$ ($v_i$) in each decomposition formalism, we can replace the differential terms in the right hand sides of the force equations  \eqref{force1}, and \eqref{force2}, using the above two equations respectively. This will make our 3-force equations consistent with the covariant equations of motion by including the electromagnetic field tensor, and consequently the Lorentz 3-force (refer to equations \eqref{force1-2}, and \eqref{force2-1}).\\
As pointed out, decomposition formalisms are based on the projection of geometric quantities along the 4-velocity of a given observer \cite{MEN}. In the next two sections we will show how one can apply the above definition of the 3-force in the two well known decomposition formalisms, namely {\it threading}, and {\it slicing}, to charge carriers in conductors and superconductors, to arrive at the generalized Schiff-Barnhill, and Meissner effects in stationary gravitational fields as measured by the observers corresponding to these decomposition formalisms.\\
To do so we need to write  the components of the 4D objects (e.g the Maxwell tensor, and 4-velocity) in terms of the 3D objects in the corresponding decomposition formalism. These 3D objects include  particle's  3-velocity, as well as the electromagnetic and gravitoelectromagnetic (GEM) fields.
We will also use the canonical momentum obtained from the Lagrangian \eqref{Lag}, in terms of the above-mentioned  3D objects to arrive at the Meissner effect in the presence of a stationary gravitational field.\\
The consistency of the force  definition \eqref{FORCE} will show itself in our derivation of the total 3-force, both the gravitational and electromagnetic (Lorentz) forces, acting on a charged particle in the slicing decomposition. Indeed, as an interesting byproduct, we find, in a  systematic way, not only the same {\it gravitational} 3-force as was found in the membrane paradigm of Thorne et.al \cite{Thorne}, but also the exact {\it Lorentz} 3-force acting on the charged particle. It will be noted that the Lorentz part of our 3-force includes the spacetime metric components, whereas in \cite{Thorne}, it was introduced based on the local consideration, and does not include the metric components.
\section{Force equation in the threading decomposition formalism}
In the threading decomposition formalism, one employs the radar signals sent between two nearby fundamental observers to define spatial distances  and time intervals \cite{LL,LBNZ}. This formalism has been employed to derive gravitational analogs of some well known electromagnetic ­effects, and also to find and interpret exact solutions of the ­Einstein field equations \cite{Nouriz}-\cite{Fil}. Also it is used to study  gravitational ­lensing. In this decomposition the spacetime metric is written in the following decomposed form,
\begin{align}\label{metrth}
ds^2= {d\tau_t}^2- dl_t^2 =  g_{00}(dx^{0}-g_{i}dx^{i})^{2} -\gamma_{ij}dx^{i}dx^{j}
\end{align}
in which $g_i = -\frac{g_{0i}}{g_{00}}$ is the so called {\it gravitomagnetic} potential, and
\begin{gather}
{d\tau}_t^2={g_{00}}(dx^{0}-g_{i}dx^{i})^{2}\label{dt} \\
dl_t^2 = {\gamma}_{ij}dx^{i}dx^{j} =\left( -g_{ij} + g_{00} g_{i}g_{j} \right)dx^{i}dx^{j}.\label{dl}
\end{gather}
are the {\it synchronized} proper time, and the metric of the 3-dimensional (quotient) space respectively. From the previous section the covariant 3-velocity in this decomposition is given by
\begin{equation}\label{4-2}
{v_t}^i = \frac{dx^i} {\sqrt{g_{00}}(dx^0 - g_i dx^i)},
\end{equation}
and
\begin{equation}\label{4}
{ds}  = \sqrt{1 - {{v_t}^2}} ~d\tau_t \equiv \frac{1}{\upgamma_t} ~d\tau_t
\end{equation}
in which $\upgamma_t = \frac{1}{\sqrt{1- {{v_t}^2}}}$ is the Lorentz factor in the same decomposition. \\
Applying this decomposition formalism to the force equation \eqref{force1}, we need the following relation between the components of 3-velocity and 4-velocity,
\begin{eqnarray}\label{4u}
u^i= \upgamma_t {{v_t}^i} \;\;\; ; \;\;\; u^0 = \upgamma_t \left( \frac{1}{\sqrt{g_{00}}}+g_{i}{{v_t}^i}\right)\\
u_{i} = -\upgamma_t({{v_t}_i} + g_i \sqrt{g_{00}}) \;\;\; ; \;\;\;
u_{0} = \upgamma_t\sqrt{g_{00}} \label{4u1},
\end{eqnarray}
Now using the first of the  equations \eqref{4u}, we can write the force equation \eqref{force1}, as follows
\begin{equation}
{f_t}^{i} = \frac{m} {\upgamma_t} \left(\frac{d u^{i}}{ds} + {\upgamma_t ^2} \lambda^i_{\;j k} v^{j}v^{k}\right), \label{force1-1}
\end{equation}
and substituting from \eqref{eom} for $\frac{d u^{i}}{ds}$, we end up with
\begin{equation}
{f_t}^{i} = \frac{m} {\upgamma_t} \left(\frac{q}{m} F^{i\mu}u_{\mu} - \Gamma^i_{\mu\nu} u^\mu u^\nu + {\upgamma_t ^2} \lambda^i_{\;j k} v^{j}v^{k}\right), \label{force1-2}
\end{equation}
in which the first term gives the Lorentz force, and the next two terms the gravitational force acting on the charged particle.
Starting from the Lorentz part of the force equation i.e,
\begin{equation}\label{FL1}
{f}^{i}_L = \frac{q}{\upgamma_t} F^{i\mu}u_{\mu}= \frac{q}{\upgamma_t}(F^{i0}u_{0} + F^{ij}u_{j}).
\end{equation}
we employ the relations \eqref{4u}, and the following definitions of the Maxwell tensor in terms of the electromagnetic fields \cite{MEN}
\begin{equation}\label{DHT}
D^{i}=-\sqrt{g_{00}}F^{0i} \;\; ; \;\;  H_{i}=-\frac{1}{2}\sqrt{g_{00}} \eta_{ijk}F^{jk}\;\;  \left( F^{ij}=-\frac{1}{\sqrt{g_{00}}} \eta^{ijk} H_k \right).
\end{equation}
to find
\begin{align}
{f}^{i}_L &= \frac{q}{\upgamma_t}\left( {\upgamma_t}{D^i} + \frac{\upgamma_t}{\sqrt{g_{00}}} \eta^{ijk} H_k ({{v_t}_j} + g_j \sqrt{g_{00}}) \right) \nonumber\\
&={q} (D^i + \frac{1}{\sqrt{g_{00}}} \eta^{ijk} {v_t}_j H_k +
\eta^{ijk} g_j H_k ).
\end{align}
Or in vector notation
\begin{equation}\label{FL}
{{\boldsymbol f}}_L = {q} ({\bf D} + {\boldsymbol g} \times {\bf H}) + \frac{q} {\sqrt{g_{00}}} ({{\boldsymbol{v}}_t} \times {\bf H}).
\end{equation}
The gravitational part of the force is given by
\begin{equation}\label{FG}
{f}^{i}_G = \frac{m}{\gamma}(\lambda^i_{\;j k} u^{j}u^{k} - \Gamma^i_{00} (u^0)^2 - 2\Gamma^i_{0j} u^0 u^j - \Gamma^i_{jk} u^j u^k)
\end{equation}
From the above equation we need  the following components of the Christoffel symbol written in terms of the 3D quantities defined in this formalism,
\begin{align}\label{5}
\Gamma^{i}_{00}&=\frac{1}{2} \nabla^{i}g_{00},\nonumber\\
\Gamma^{i}_{0j}&=\frac{g_{00}}{2} (\nabla_{j}g^{i}- \nabla^{i}g_{j})- \frac{1}{2}g_{j}\nabla^{i}g_{00},\nonumber\\
\Gamma^{i}_{jk}&= \lambda^{i}_{jk}-\frac{g_{00}}{2}[g_{j}(\nabla_{k}g^{i}- \nabla^{i}g_{k})+g_{k}(\nabla_{j}g^{i}- \nabla^{i}g_{j})]+\frac{1}{2}g_{j}g_{k}\nabla^{i}g_{00}.
\end{align}
in which $\nabla$ denotes the 3D covariant derivative in the 3D space represented by the metric $\gamma_{ij}$ in this decomposition, and $\lambda^{i}_{jk}$ is given by \eqref{Lamb}.
Substituting the above equations, along with the relations \eqref{4u} in  equation \eqref{FG}, and after a long but straightforward calculation we end up with the following form for the gravitational force
\begin{align}
{f_G}^{i}&= \upgamma_t {m}\left\lbrace
-\nabla^{i}\ln{\sqrt{g_{00}}}+{\sqrt{g_{00}}}({\boldsymbol{v}_t}\times(\boldsymbol{\nabla}\times{\boldsymbol g}))^{i}
\right\rbrace.
\end{align}
In vector notation, and after using the definitions of the gravitoelectric, and gravitomagnetic fields in the threading formalism, namely,
\begin{equation}\label{GEMT}
\boldsymbol{E}_g = -\boldsymbol{\nabla}\ln \sqrt{g_{00}} \;\;  ; \;\; \boldsymbol{B}_g = {\bf \nabla} \times {\boldsymbol g},
\end{equation}
it transforms to
\begin{align}\label{GT}
{\boldsymbol f}_G &= \upgamma_t  {m} (
\boldsymbol{E}_g + {\sqrt{g_{00}}}{\boldsymbol{v}_t}\times\boldsymbol{B}_g).
\end{align}
Summing the two contributions \eqref{FL}, and \eqref{GT},  we end up with the following elegant 3-force equation acting on a charged particle in gravitational and electromagnetic fields,
\begin{align}\label{FTT}
{\boldsymbol f}_t &= \upgamma_t {m}\left\lbrace
{\bf E}_g+\sqrt{g_{00}} {{\boldsymbol{v}_t}}\times {\bf B}_{g}
\right\rbrace + \frac{q}{\sqrt{g_{00}}} \left\lbrace  \sqrt{g_{00}} ({\bf D} + {\boldsymbol g} \times {\bf H}) + {{\boldsymbol{v}_t}} \times {\bf H} \right\rbrace \cr
&= \upgamma_t {m}\left\lbrace
{\bf E}_g+\sqrt{g_{00}} {{\boldsymbol{v}_t}}\times {\bf B}_{g}
\right\rbrace + \frac{q}{\sqrt{g_{00}}} \left\lbrace  {\bf E}+\sqrt{g_{00}} {{\boldsymbol{v}_t}} \times ({\bf B}-{\boldsymbol g}\times{\bf E})\right\rbrace,
\end{align}
where in the second line we have substituted for $\bf D$ and $\bf H$ in terms of $\bf E$ and $\bf B$ from the following gravitationally-induced constitutive equations \cite{MEN},
\begin{equation}\label{ce}
{\bf D}=\frac{{\bf E}}{\sqrt{g_{00}}} + {\bf H} \times {\boldsymbol g}~~~~;~~~~ {\bf B}=\frac{\bf H}{\sqrt{g_{00}}} - {\bf E} \times {\boldsymbol g}.
\end{equation}
in the threading decomposition formalism.
\section{Force equation in the slicing decomposition formalism}
In this section we calculate the force acting on a charged particle  in  gravitational and electromagnetic fields by employing the slicing decomposition formalism. From an observer's point of view this is a measurement made by a fiducial observer, one not fixed with respect to distant observers. It is expected that, as in  the case of threading decomposition, the metric components appear in the Lorentz part of the force, something that has not been appeared in previous studies due to either local or weak field limit considerations \cite{Thorne}.
In the slicing decomposition, we have the following well known form of the line element,
\begin{equation}\label{ms}
ds^2={d\tau}_S^2- dl_S^2 = N^2 (dx^0)^2 - \gamma_{ij} (N^i dx^0 + dx^i)(N^j dx^0 + dx^j),
\end{equation}
in which $N = \frac{1}{\sqrt{g^{00}}}$, and  ${N^i}= -{N^2} g^{0i}$ are the so called lapse function, and shift vector respectively. From Eqs. \eqref{nf3} and \eqref{nf4} applied to  this formalism  we have,
\begin{align}
{v_s}^i &= ( \frac{dx^i}{N dx^0} + \frac{N^i}{N}) \label{rela}.
\end{align}
and
\begin{equation}\label{dss}
 ds = \sqrt{1-{{v_s}^2}} ~d\tau_s \equiv \frac{1}{\upgamma_s} ~d\tau_s
\end{equation}
with ${v_s}_i = \gamma_{ij}{v_s}^j$, and $\upgamma_s = \frac{1}{\sqrt{1-{v_{s}^2}}}$. Using the above relations, we can find the time and spatial components of the 4-velocity $u_\mu = \frac{dx^\mu}{ds}$ in terms of the above 3-velocity and metric components as follows,
\begin{align}\label{velo0}
u_{i} = -\upgamma_s{v_{s_i}};\;\;\; ; \;\;\;
u_{0} = \upgamma_s(N - {N^{i}v_{s_i}})\\
u^{0} = \frac{\upgamma_s}{N}\;\;\;\; ; \;\;\;\;\;\;
u^{i} =  \upgamma_s ({v_{s}^{i}} - \frac{N^{i}}{N}).\label{velo}
\end{align}
In this decomposition things are more simpler if we use the lower- indexed force equation \eqref{force2}, in which by using the first equation in \eqref{velo0} it transforms to
\begin{equation}
{f_s}_{i} = \frac{m} {\upgamma_s} \left(-\frac{d u_{i}}{ds} + \frac{1}{2}\upgamma_s^2 \gamma_{j k, i} v^{j}v^{k}\right), \label{force2-1}
\end{equation}
On the other hand from the equation of motion \eqref{eoml},  we can replace for $\frac{d u_{i}}{ds}$ to end up with,
\begin{equation}
{f_s}_{i} = - \frac{m} {\upgamma_s} \left(\frac{1}{2}g_{\mu\nu,i} u^\mu u^\nu + \frac{q}{m} F_{i\mu}u^{\mu} - \frac{1}{2}\upgamma_s^2 \gamma_{j k, i} v^{j}v^{k}\right), \label{force2-1}
\end{equation}
As in the threading case we start from the Lorentz part of the force
which could be written as
\begin{equation}\label{FL1}
{{f}_L}_i = - \frac{q}{\upgamma_s} F_{i\mu}u^{\mu}= - \frac{q}{\upgamma_s}(F_{i0}u^{0} + F_{ij}u^{j}).
\end{equation}
Employing the relations \eqref{velo0}, and \eqref{velo}, along with the following definitions of the Maxwell tensor in terms of the electromagnetic fields \cite{MEN},
\begin{equation}\label{ems}
E_{i}=F_{0i},~~~~ B^{i}=-\frac{1}{2}\eta^{ijk} B_{jk},
\end{equation}
in the above equation we find
\begin{align}
{{f}_L}_i  = - \frac{q}{\upgamma_s}\left\lbrace - E_i \frac{\upgamma_s}{N}  -\upgamma_s \eta_{ijk} B^{k}({v_{s}^{j}}-\frac{N^{j}}{N})\right\rbrace.
\end{align}
or in vector notation
\begin{align}\label{FLS}
{{\boldsymbol f}}_L = q ~\left( \frac{\bf E}{N} + {\boldsymbol{v}_s}\times \boldsymbol{B} -  \frac{\boldsymbol{N}\times \boldsymbol{B}}{N}\right).
\end{align}
The gravitational part is given by
\begin{align}
{f_G}_{i} = - \frac{m} {\upgamma_s} \left(\frac{1}{2}g_{\mu\nu, i} u^\mu u^\nu - \frac{1}{2}\upgamma_s^2 \gamma_{j k, i} v^{j}v^{k}\right)\;\;\;\;\;\;\;\;\;\;\;\;\;\;\;\;\;\;\;\;\;\;\;\;\;\;\;\;\;\;\; \nonumber \\
 = - \frac{m} {2 \upgamma_s}\left(g_{00, i} (u^0)^2 + 2 g_{0j, i} u^0 u^j + g_{jk,i} u^j u^k- \upgamma_s^2 \gamma_{j k, i} v^{j}v^{k}\right),
\end{align}
which upon substituting for the metric components from \eqref{ms}, and the 3-velocity components from  \eqref{velo}, after a long but straightforward calculation we end up with
\begin{align}
{f_G}_i =\upgamma_s {m} \left\lbrace -\frac{1}{2N^2}\partial_{i} N^2+\frac{\partial_{i}N_{j}}{N} {v_{s}^{j}}\right\rbrace.
\end{align}
In vector form, it can be expressed as
\begin{align}\label{FGS}
{\boldsymbol f}_G &=\upgamma_s {m} \left\lbrace -\boldsymbol{\nabla}\ln N + {{\boldsymbol{v}_t}_s}\times\left( \frac{{\bf \nabla}\times {\bf N}}{N}\right) \right\rbrace  \cr
&= \upgamma_s {m} \left\lbrace \boldsymbol{E_{g}} +\frac{{\boldsymbol{v}_s}}{N}\times\boldsymbol{B_{g}}
\right\rbrace,
\end{align}
where in the second line we have used the following definitions of the gravitoelectric, and gravitomagnetic fields in the slicing formalism \cite{MEN, Thorne},
\begin{equation}\label{GEMS}
\boldsymbol{E}_g = -\boldsymbol{\nabla}\ln N \;\;  ; \;\; \boldsymbol{B}_g = {\bf \nabla}\times {\bf N}.
\end{equation}
Now adding the Lorentz \eqref{FLS} and gravitational \eqref{FGS} parts of the force we finally have the illuminating relation
\begin{align}\label{fss}
{\boldsymbol f}_s &=\upgamma_s {m} \left\lbrace \boldsymbol{E_{g}} +\frac{{\boldsymbol{v}_s}}{N}\times\boldsymbol{B_{g}}\right\rbrace
+q ~\left( \frac{\bf E}{N} + {\boldsymbol{v}_s}\times \boldsymbol{B} -  \frac{\boldsymbol{N}\times \boldsymbol{B}}{N}\right) \cr
&= \upgamma_s {m} \left\lbrace \boldsymbol{E}_g + \frac{\boldsymbol{v}_s}{N}\times \boldsymbol{B}_g \right\rbrace +
 q \left\lbrace \boldsymbol{D} + {\boldsymbol{v}_s}\times \boldsymbol{B} \right\rbrace .
\end{align}
where in the second line we have used the  gravitationally-induced constitutive equations in the slicing decomposition formalism \cite{MEN}, namely,
\begin{equation}\label{ces}
{\bf D}=\frac{{\bf E}}{N} + \frac{1}{N}{\bf B} \times {\bf N}~~~~;~~~~ {\bf B}=\frac{\bf H}{N} - \frac{1}{N}{\bf D} \times {\bf N}.
\end{equation}
to write the electromagnetic part of the force in terms of $\bf D$, and $\bf B$.\\
The Lorentz part of the force equation \eqref{FLS}, which we  obtained through a systematic approach,  is different from that mentioned in the membrane paradigm approach in \cite{Thorne}. There the authors, based on local measurement, take the Lorentz force acting on the particle (in the case of kerr metric geometry) to have the same form as that in the flat spacetime. Indeed, considering the equivalence principle, one should be more cautious with this local consideration of electromagnetic force on a particle in a gravitational field, while  simultaneously find the gravitational force \eqref{FGS} acting on the same particle. In other words our result is more general in the sense that it explicitly shows you find different results in the presence, and in the absence of the gravitational field.
\section{Dynamics of superconducting fluid : Generalized Schiff-Barnhill and Meissner effects}
In his phenomenological treatment of superconductivity, London considered the motion of the so called superelectrons in a superconductor as the {\it rotational} motion of a hydrodynamical non-viscous electronic fluid in an electromagnetic field, taking into account the Meissner effect \cite{London}. In this treatment the so called London equations originate from the force acting on superelectrons (Cooper pairs in BCS theory) in an electromagnetic field \cite{Feynman}.  Now if the superconductor is also placed in a gravitational field, one can take equations \eqref{FTT}, and \eqref{fss}  as the force acting on free superelctrons in the superconductor as measured by the two sets of observers corresponding to the two decomposition formalism. In the following subsections we follow this scenario by starting with the simple case of a static spacetime, where the two decomposition formalisms coincide. Then we consider the case of a (type-I) superconductor in a stationary spacetime where the two decomposition formalisms, and their corresponding observers are different, leading to observer-dependent results.
\subsection{Schiff-Barnhill effect in static spacetimes}
Applying the force equations \eqref{FTT} and \eqref{fss} to the case of a charged particle in a static spacetime (where ${\bf B}_g = 0$, and  ${\boldsymbol g} = {\bf N}=0$), the two observers and their corresponding decompositions will be the same. Indeed in this case both proper times reduce to,
\begin{equation}
d\tau = {\sqrt{g_{00}}} dx^0
\end{equation}
so that the 3-velocity is given by
\begin{equation}
v^i =  \frac{dx^i}{d\tau} \;\;\;\ ; \;\;\; v^2 = -\frac{g_{ij}dx^i dx^j}{d\tau^2}
\end{equation}
leading to fact that the two $\upgamma$-factors are the same i.e
\begin{equation}
\upgamma_s = \upgamma_t = \frac{1}{\sqrt{1-v^2/c^2}}.
\end{equation}
Taking the above facts into account, in the absence of any magnetic field ($\bf B = 0$), we end up  with the following force equation in both formalisms,
\begin{equation}
{\boldsymbol f } = \frac{{}^3D{\bf p}}{d\tau} = q {\bf E} + {m} \gamma \sqrt{g_{00}}  {\bf E}_g.
\end{equation}
Now if we apply the above force equation to the free charges (electrons) in a conductor, and require the electrostatic equilibrium (the stationary condition), where the force acting on the charged particle vanishes, we have
\begin{equation}
{\bf E} = -  \frac{m}{q} \upgamma \sqrt{g_{00}} {\bf E}_g.
\end{equation}
This is the  the generalized Schiff-Barnhill effect for a (super)conductor in a  static spacetime, without any weak field or low velocity limit, including the metric component $g_{00}$, and the
$\gamma$-factor. Obviously in the weak field and low velocity limit, it reduces to \eqref{F1}.
\subsection{Generalized Schiff-Barnhill and Meissner effects in threading formalism}
Now taking the equation \eqref{FTT} as the force acting on superelectrons (Cooper pairs $q=2e$) in a superconductor placed in  gravitational and electromagnetic fields, its terms could be rearranged as follows;
\begin{align}\label{FT}
{\boldsymbol f}_t = \left\lbrace  {m}\upgamma_t
{\bf E}_g+\frac{q}{\sqrt{g_{00}}}{\bf E}
\right\rbrace +\left\lbrace {m}\upgamma_t \sqrt{g_{00}}{{\boldsymbol{v}_t}} \times {\bf B}_{g} +{q} {{\boldsymbol{v}_t}} \times ({\bf B}-{\boldsymbol g}\times{\bf E})\right\rbrace.
\end{align}
where the force acting on superelectrons is divided into two parts, one independent of the particle's velocity and the other one dependent on it. Now in the presence of the gravitational field we require the stationary condition in the bulk of the superconductor, i.e ${\bf F}_t = 0$. This in general requires vanishing of both terms separately, leading to the following two equations \footnote{It is noted that, as in the original case, $\boldsymbol v$ does not necessarily vanish in the bulk.},
\begin{equation}\label{MSBE}
{m}\upgamma_t
{\bf E}_g+\frac{q}{\sqrt{g_{00}}}{\bf E}=0
\end{equation}
and
\begin{equation}\label{MME}
{\bf B} +  \frac{m}{q} \upgamma_t~{\sqrt{g_{00}}}~ {\bf B}_g - {\boldsymbol g} \times {\bf E} = 0
\end{equation}
which are respectively the generalized Schiff-Barnhill, and Meissner effects in the presence of a stationary gravitational field. To make sure that the above stationary condition is not an ad hoc one, in what follows we show the consistency of our approach by finding  the same result, \eqref{MME}, from the generalized second London  equation. To find the second London  equation in the presence of a gravitational field lets  recall the original second London equation \cite{London},
\begin{equation}
 {\boldsymbol \nabla} \times {\boldsymbol v} = -\frac{q}{m} {\bf B},
\end{equation}
which describes the motion of the charge carriers in the superconductor as an ideal fluid in a magnetic field. Indeed instead of the velocity field, one can write the above equation in terms of the supercurrent ${\bf j}_s = qn_s{\boldsymbol v}$, where $n_s$ is the number density of superelectrons. On the other hand  it is noted that the above equation is nothing, but the vanishing of the curl of the canonical (dynamical) momentum ${\bf P} = {\bf p}+ q {\bf A}$. Indeed  one can define a generalized vorticity ${\boldsymbol \omega} = \frac{1}{m} {\boldsymbol \nabla} \times {\boldsymbol P} = {\boldsymbol \nabla} \times ({\boldsymbol v} + \frac{q}{m} {\bf A})$, so that the above relation is equivalent to the vanishing of the $\boldsymbol \omega $, either in the interior of a plasma reaching the equilibrium states of perfect diamagnetism \cite{Mahajan}, or in the body of a type-I superconductor beyond the London penetration depth \cite{Hirsch}.
\footnote{In the quantum mechanical treatment of superconductivity in the BCS theory, by assigning a wave function to a Cooper pair in the ground state, one can show that the canonical momentum assigned to the electronic fluid is given by the gradient of the phase of the wave function, leading to its irrotational character \cite{BCS, Feynman}. Indeed one can show that the same reasoning works in a quantum mechanical treatment of  Cooper pairs in the presence of a stationary gravitational field \cite{MN}.}. By taking into account the Meissner effect inside the superconductor,  beyond the penetration depth (i.e ${\bf B} =0$), we end up with the irrotational character of velocity field (supercurrent), ${\boldsymbol \nabla} \times {\boldsymbol v} = 0$. \\
To obtain the modification of this equation in the presence of a stationary gravitational field in any decomposition formalism, first we need to calculate the spatial components of the particle's canonical 4-momentum. Employing the Lagrangian \eqref{Lag}, one can show that in terms of the decomposed 3D metric components in the threading formalism, and in vector notation, it is given by (refer to Appendix),
\begin{equation}
{\bf P} = m\upgamma_t\left( \sqrt{g_{00}} {\boldsymbol g} + {\boldsymbol{v}_t} \right) - q g_{00} {\boldsymbol g}
\left(A^0 + {\boldsymbol g}.{\bf A} \right) + {q} {\bf A}.
\end{equation}
Now  setting  its curl equal to zero, and following London to ignore nonlinear terms, we end up with the following equation (refer to appendix A for detailed calculation)
\begin{equation}\label{SLET}
 \upgamma_t {\boldsymbol \nabla} \times {\boldsymbol v}_t = -\left( \upgamma_t {\bf B}_g + \frac{q}{m \sqrt{g_{00}}} {\bf B} +  \upgamma_t ({\boldsymbol g} \times {\bf E}_g)\right)
\end{equation}
which is London's second equation in the presence of a stationary gravitational field. Indeed, in the absence of the gravitational field, it reduces to the relativistic version of London's second equation
${\boldsymbol \nabla} \times (\upgamma_t {\boldsymbol v}_t) =
-\frac{q}{m} {\bf B}$
given by London himself (See Eq. (22) in \cite{London}). To find the generalized Meissner effect, we take the ansatz that the presence of a stationary gravitational field will not change the irrotational character of the velocity field inside the superconductor, so that the left hand side of the above equation vanishes leading to
\begin{equation}\label{L2}
  \upgamma_t {\bf B}_g + \frac{q}{m \sqrt{g_{00}}} {\bf B} +  \upgamma_t ({\boldsymbol g} \times {\bf E}_g) = 0
\end{equation}
The above equation is the Meissner effect in the presence of a stationary gravitational field.
To show that our approach is consistent, and the above equation is equivalent to our previous result \eqref{MME}, we use the generalized Schiff-Barnhill effect \eqref{MSBE}, i.e,
\begin{equation}
{\bf E}_g = - \frac{q}{m\upgamma_t~{\sqrt{g_{00}}} }{\bf E}  \label{E1}
\end{equation}
to replace  ${\bf E_g}$ with ${\bf E}$ in equation \eqref{L2},
leading to \eqref{MME}. Writing Eqs. \eqref{MSBE} and \eqref{MME} in the following forms
\begin{equation}
{\bf E} = - \frac{m}{q} \upgamma_t \sqrt{g_{00}}{\bf E}_g = 0,
\end{equation}
and
\begin{equation}\label{IMF}
\boldsymbol{B} = - \frac{m\upgamma_t}{q} \sqrt{g_{00}} \boldsymbol{B_{g}}
+(\boldsymbol{g \times E}).
\end{equation}
which are the  the generalizations of the Schiff-Barnhill, and meissner effects to the high velocity and full field regime, as measured by a fundamental observer \cite{MEN}. Indeed as measured by such an observer there are  induced electric and magnetic fields inside the superconductor given by the above values. Obviously they both reduce to the usual effects in the absence of a gravitational field.\\
It is noted that compared to the previous results including that of De Witt , which was in the weak field limit, now there is a new term contributing to the gravitationally-induced magnetic field inside a superconductor, namely ${\boldsymbol g} \times {\bf E}$. This was somewhat expected from the gravitationally-induced constitutive equations \cite{MEN} as measured by the fundamental observers corresponding to the  threading decomposition formalism. \\
So in this decomposition formalism  it is not the magnetic field plus gravitomagnetic field which vanishes inside a super conductor, but there is an additional induced term which is  basically due to a magnetoelectric effect, originating from the time-asymmetry in the underlying stationary spacetime \cite{MEN, Fiebig}.
\subsection{Generalized Schiff-Barnhill and Meissner effects in slicing formalism}
As in the threading case, rearranging terms in the force equation \eqref{fss}, we have
\begin{align}\label{TFS}
{\boldsymbol f}_s = \left( {m} \upgamma_s  \boldsymbol{E}_g +  q \boldsymbol{D} \right)
+ \left( q {\boldsymbol{v}_s}\times \boldsymbol{B}+ {m } \frac{\upgamma_s}{N} {\boldsymbol{v}_s}\times \boldsymbol{B}_g  \right).
\end{align}
As in the threading case, we now apply the stationary condition
$\boldsymbol{F}_s = 0$ and we find the following two relations
\begin{equation}\label{SBES}
{m} \upgamma_s  \boldsymbol{E}_g +  q \boldsymbol{D} = {m} \upgamma_s  \boldsymbol{E}_g + q (\frac{{\bf E}}{N} + \frac{1}{N}{\bf B} \times {\bf N}) = 0
\end{equation}
and
\begin{equation}\label{MES}
q \boldsymbol{B}+ {m} \frac{\upgamma_s}{N} \boldsymbol{B}_g = 0
\end{equation}
as the generalized versions of the Schiff-Barnhill and Meissner effects respectively, due to the presence of a stationary gravitational field.
Now as in the previous subsection we try to find the generalized Meissner effect from the second London equation in a stationary spacetime decomposed in slicing formalism.
To this end first we find the canonical momentum in a stationary background in the slicing formalism as follows (refer to appendix A),
\begin{equation}
 \boldsymbol{P} = -{m} \upgamma_s (\boldsymbol{v}_s + \frac {\boldsymbol{N}}{N}) - q  \boldsymbol{A}.
\end{equation}
Now as in the threading formalism,  setting the curl of the above relation equal to zero while ignoring nonlinear terms, we are led to (refer to appendix A)
\begin{equation}\label{SL2}
-\upgamma_s {\boldsymbol \nabla} \times {\boldsymbol v}_s = \upgamma_s \boldsymbol{B}_g + \frac{q}{m N}\boldsymbol{H} +
\upgamma_s ({\boldsymbol{E}_g}\times \frac{\boldsymbol{N}}{N}),
\end{equation}
which is the second London equation in slicing formalism. This in turn by taking the ansatz that the presence of a stationary gravitational field will not change the irrotational character of the velocity field inside the superconductor, leads to the generalized Meissner effect,
\begin{equation}\label{MMES}
\boldsymbol{B}_g + \frac{q}{m N \upgamma_s }\boldsymbol{H} +
 ({\boldsymbol{E}_g}\times\frac{\boldsymbol{N}}{N}) = 0.
\end{equation}
To show that the above result is equivalent to Eq. \eqref{MES}, first we use the generalized Schiff-Barnhill effect \eqref{SBES}, to replace ${\bf E}_g$ with $\bf D$ so that it reduces to
\begin{equation}
\boldsymbol{B}_g + \frac{q}{m N \upgamma_s }\boldsymbol{H} -\frac{q}{m\upgamma_s}({\boldsymbol{D}}\times\frac{\boldsymbol{N}}{N}) = 0.
\end{equation}
Now using the gravitationally-induced constitutive equations \eqref{ces}, the above equation reduces to \eqref{MES} . So writing the two equations \eqref{SBES} and \eqref{MES} in the following forms,
\begin{equation}\label{SBES1}
{{\bf E}} = - \frac{ N {m} \upgamma_s}{q}  \boldsymbol{E}_g - {\bf B} \times {\bf N}
\end{equation}
and
\begin{equation}\label{MES1}
\boldsymbol{B} = - \frac{m \upgamma_s}{q N}\boldsymbol{B}_g.
\end{equation}
they represent the gravitationally-induced electric and  magnetic fields inside a superconductor as measured by a fiducial observer in a stationary spacetime.
\subsection{Comparison of the results in the two decomposition formalisms}
It is illuminating to compare the above results, in the sense of changes made to the Schiff-Barnhill, and Meissner effects in the presence of a gravitational field, as measured by the two sets of observers corresponding to the two decomposition formalisms.\\
{\bf I}-It is noted that in both formalisms metric components, as well as  $\gamma$-factors enter both effects such that they reduce to the previous results \cite{Schiff}-\cite{DeWitt}, in the weak field, and low velocity limit.\\
{\bf II}-Looking at the generalized Schiff-Barnhill effect, it is noted that in  the threading formalism, compared to the weak field limit, there are no extra terms. On the other hand in the slicing formalism there is an extra term which is proportional to $\boldsymbol{N}\times \boldsymbol{B}$. \\
{\bf III}-In the case of generalized Meissner effect, it is noticed that there is an extra term in the threading formalism which is proportional to $\boldsymbol{g}\times \boldsymbol{E}$, while there are no extra terms in the slicing formalism. \\
In other words the presence of the gravitationally-induced magnetoelectric effect reflects itself in the modifications of the Schiff-Barnhill, and  Meissner effects in the slicing and threading decomposition formalisms respectively. Obviously this could be traced back to the nature of the two sets of observers, fundamental and fiducial, corresponding to the two decomposition formalisms.
\section{Summary and conclusions}\label{conclusions}
Previous studies on conductors and superconductors in gravitational and electromagnetic fields, are mostly restricted to the limit of weak fields and low velocities. In the present study we have considered the gravitationally-induced electric and magnetic fields in conductors and superconductors, and  the related phenomena, without employing these restrictions.
We followed London's phenomenological approach to superconductivity which is based on the treatment of the motion of charge carriers in the context of the hydrodynamics of an electrically charged ideal fluid, taking into account the Meissner effect \cite{Feynman}. Employing the same approach to study the electrodynamics of superconductors in stationary spacetimes,  we introduced the 3-force acting on a charged particle in gravitational and electromagnetic fields in a given spacetime decomposition formalism. Since the spacetime decomposition formalisms are based on projection of geometric quantities along the 4-velocity of a given observer, and consequently allow for definition of other 3D physical quantities, this approach highlights the role of observers in such phenomena. Indeed to the best of our knowledge, this is the first study in which the role of observers is taken into account in the electrodynamics of superconductors in curved backgrounds. To show the role of observers more explicitly, we have employed the well-known threading and slicing decomposition formalisms to the stationary spacetimes, corresponding to the fundamental, and fiducial observers respectively. Employing the stationary condition on charge carriers in a superconducting  state, we arrived at the generalized Schiff-Barnhill and Meissner effects under a single formulation. Our study showed that compared to the weak field and low velocity limit, extra terms contribute to these effects arising from the gravitationally induced constitutive equations in each decomposition formalism \cite{MEN}.\\
Finally it is noted that compared to the well established quantum mechanical treatment of superconductivity in the BCS theory, London's approach works because the quantum many-body state of a type-I superconductor looks like a classical charged superfluid on macroscopic scales.
Therefore it is expected that one should be able to follow the same study in the context of quantum mechanics in a stationary spacetime by applying the covariant Klein-Gordon equation to the Cooper pairs. Obviously in this case we also need to take into account the role of observers by applying different spacetime decomposition formalism to the underlying spacetime \cite{MN}.
\section *{Acknowledgments}
M. N-Z and F. M  would like to thank University of Tehran for supporting this project under the grants provided by the research council.
\appendix
\section{Dynamical momentum and the second London equation in stationary spacetimes}
In this appendix we present the detailed derivation of  the canonical momentum, and the second London equation used in the text for the two decomposition formalisms. Applying the  Euler–Lagrange prescription to  the Lagrangian \eqref{Lag} we obtain the canonical momentum as follows
\begin{equation}
P_{\mu}=-mg_{\mu\nu}u^{\nu}-q g_{\mu\nu} A^{\nu}.\label{gma1}
\end{equation}
On the other hand in any decomposition formalism, the curl of a 3-vector $A^{i}$ is defined by
\begin{equation}
(\boldsymbol{\nabla}\times \boldsymbol{A})^{i}=\frac{1}{2 \sqrt{\gamma}} \epsilon^{ijk}\left(  \nabla_{j} A_{k} - \nabla_{k} A_{j}\right),\label{gma3}
\end{equation}
in which ${\gamma}$ is the determinant of the spatial 3-metric in the corresponding decomposition formalism. In what  follows, we will
use the above two relations to find the London second equation in both the threading, and slicing decomposition formalisms, namely equations  \eqref{SLET}, and \eqref{SL2}.
\subsection*{Threading formalism}
Starting from the covariant spatial components of the equation \eqref{gma1}
\begin{align}\label{sgma1}
P_{i}& =-m g_{i \nu}u^{\nu}-q g_{i \nu} A^{\nu} \nonumber \\
& = -m g_{i0}u^{0} -mg_{ij}u^{j} - q g_{i0} A^{0}- q g_{ij} A^{j}
\end{align}
and substituting for the metric, and 4-velocity components from \eqref{dl}, and \eqref{4u1} respectively, we end up with
\begin{align}
P_{i}& = g_{00}g_{i}\left\lbrace  m \upgamma_t( \frac{1}{\sqrt{g_{00}}}+\boldsymbol{g.v_t})+q A^{0}\right\rbrace +(\gamma_{im}-g_{00}g_{i}g_{m})(\upgamma_tv^{m}_{t}+qA^{m}) \nonumber \\
& = m \upgamma_t \left\lbrace \sqrt{g_{00}} g_{i}+v_{ti}\right\rbrace +q g_{00}g_{i}\left\lbrace A^{0}-\boldsymbol{g.A}\right\rbrace + q A_{i}. \label{GMT}
\end{align}
or in vector notation
\begin{equation}
{\bf P} = m\upgamma_t\left( \sqrt{g_{00}} {\boldsymbol g} + {\boldsymbol{v}_t} \right) - q g_{00} {\boldsymbol g}
\left(A^0 + {\boldsymbol g}.{\bf A} \right) + \frac{q}{c} {\bf A}.
\end{equation}
Now setting the curl of the above equation equal to zero we have
\begin{equation}
{\bf \nabla}\times{\bf P}= m \upgamma_t {\boldsymbol \nabla} \times  {\boldsymbol v}_t + m \upgamma_t {\bf \nabla}\times (\sqrt{g_{00}}{\boldsymbol g})+ q \boldsymbol{\nabla}\times \left\lbrace g_{00} \boldsymbol{g}(A^{0}-{\bf g.A}) \right\rbrace+ q \boldsymbol{\nabla}\times \boldsymbol{A}=0. \label{eq1}
\end{equation}
in which, following London, we have ignored nonlinear terms which include $\nabla ({\bf v}^2)$. Using the definitions of the gravitoelectric, and gravitomagnetic fields from  \eqref{GEMT}, the second term on the right-hand side of the above equation can be calculated as
\begin{equation}
{\boldsymbol \nabla} \times (\sqrt{g_{00}}{\boldsymbol g})=-\sqrt{g_{00}} ~\boldsymbol{E_{g}}\times {\boldsymbol g}+ \sqrt{g_{00}} {\bf B_{g}},\label{eq2}
\end{equation}
which upon  substitution in \eqref{eq1}, we find
\begin{equation}
m \upgamma_t {\boldsymbol \nabla} \times  {\boldsymbol v}_t + q\boldsymbol{\nabla}\times \left\lbrace \boldsymbol{A}+g_{00}~ \boldsymbol{g}~(A^{0}-\boldsymbol{g.A}) \right\rbrace+ m\upgamma_t \sqrt{g_{00}} \boldsymbol{B_{g}}-m\upgamma_t \sqrt{g_{00}}(\boldsymbol{E_{g}\times g})=0. \label{eq3}
\end{equation}
The second term in the above equation could be simplified more by using the definition of the magnetic field $\mathbf{B}$ in a curved background in threading decomposition's namely,
\begin{equation}
B^{i} = - \frac{1}{2} \eta^{ijk}\left\lbrace \nabla_{j}A_{k}-\nabla_{k} A_{j} \right\rbrace, \label{BB}
\end{equation}
As previously discussed in \cite{MEN}, we should write the covariant components of the 3-potential $A_{i}$ in terms of contravariant components of $A^{i}$ which form a 3-vector in $\Sigma_3$. So we start from $A_{i} = g_{i\mu}A^{\mu}$, and use the definitions of $g_i$ and the spatial metric $\gamma_{ij}$ in threading decomposition, to obtain
\begin{equation}
A_{i}=-g_{00}g_{i}\left\lbrace A^{0}-g_{j}A^{j}\right\rbrace - \gamma_{ij}A^{j} ,
\end{equation}
hence,  eq. \eqref{BB} is rewritten as
\begin{align}
B^{i}&=\frac{1}{2}\eta^{ijk} \left\lbrace  \nabla_{j}(g_{00}g_{k}(A^{0}-\boldsymbol{g.A})) -\nabla_{k}(g_{00}g_{j}(A^{0}-\boldsymbol{g.A}))\right\rbrace \nonumber\\
&+\frac{1}{2}\eta^{ijk}\left\lbrace \nabla_{j}(\gamma_{km}A^{m})-\nabla_{k}(\gamma_{jm}A^{m})\right\rbrace,\\
&=\eta^{ijk}( \nabla_{j}\left\lbrace g_{00}g_{k}(A^{0}-\boldsymbol{g.A})+\gamma_{km}A^{m}\right\rbrace )
\label{BB1}
\end{align}
or in vector form,
\begin{align}
	\mathbf{B}= \mathbf{\nabla}\times\left\lbrace g_{00}\mathbf{g}(A^{0}-\boldsymbol{g.A}) +\mathbf{A}\right\rbrace.
	\label{BB3}
\end{align}
Substituted in \eqref{eq3} the second term is $q\boldsymbol{B}$, and we end up with the London second equation in the following form,
\begin{equation}
 - \upgamma_t {\boldsymbol \nabla} \times {\boldsymbol v}_t = \upgamma_t {\bf B}_g + \frac{q}{m \sqrt{g_{00}}} {\bf B} +  \upgamma_t ({\boldsymbol g} \times {\bf E}_g)
\end{equation}
\subsection*{Slicing formalism}
Starting from the spatial components of the contravariant canonical momentum $P^{\mu}$, i.e
\eqref{sgma1},
\begin{equation}
P^{i}=-m u ^{i}-q  A^{i},\label{gma2}
\end{equation}
and using relations \eqref{rela},  and \eqref{velo}, in the above equation we find
\begin{align}
P^{i}=-m\upgamma_s \frac{v_s^{i}}{c}-m\upgamma_s \frac{ N^{i}}{N}-qA^{i}.\label{eq7}
\end{align}
Setting the curl of the above  generalized canonical momentum  equal to zero
\begin{equation}
(\boldsymbol{\nabla}\times\boldsymbol{P})_{k}=\eta_{kij} \nabla^{i}\left\lbrace  -m\upgamma_s {v_s^{j}}-m\upgamma_s \frac{ N^{j}}{N}-qA^{j}\right\rbrace  = 0 \label{eq80}
\end{equation}
 we have
\begin{equation}
-m\upgamma_s\boldsymbol{\nabla}\times \boldsymbol{v}_s = m\upgamma_s\boldsymbol{\nabla}\times \frac{\boldsymbol{N}}{N}+q \boldsymbol{\nabla}\times \boldsymbol{A} . \label{eq8}
\end{equation}
Here, following London, again we have ignored the nonlinear terms in velocity which include ${\bf \nabla} (\boldsymbol v^2)$.
From the definitions of the electromagnetic field in \cite{MEN}, we know
\begin{equation}
H_{i}=-\frac{N}{2}\eta_{ijk}F^{jk}=-\frac{1}{2}\eta_{ijk}N\left\lbrace \nabla^{j} A^{k}-\nabla^{k} A^{j}\right\rbrace
\end{equation}
or in vector form simplifies to
\begin{equation}
\mathbf{H}=N~ \mathbf{\nabla}\times \mathbf{A},
\end{equation}
and  when substituted in \eqref{eq8} we have,
\begin{equation}
 -m\upgamma_s\boldsymbol{\nabla}\times \boldsymbol{v}_s   = m\upgamma_s\boldsymbol{\nabla}\times \frac{\boldsymbol{N}}{N}+\frac{q}{N} \boldsymbol{H}. \label{eq9}
\end{equation}
On the other hand using the relation
\begin{equation}
\boldsymbol{\nabla}\times \frac{\boldsymbol{N}}{N}=\frac{1}{N}\boldsymbol{\nabla}\times \boldsymbol{N}+ \boldsymbol{\nabla}(\frac{1}{N})\times\boldsymbol{N}, \label{eq13}
\end{equation}
along with the definition of the gravitoelectric field in the slicing formalism  to calculate
\begin{equation}
\boldsymbol{\nabla}(\frac{1}{N})=-\frac{\boldsymbol{\nabla}N}{N^2}
=-\frac{1}{N}\boldsymbol{\nabla}(\ln N)=\frac{\boldsymbol{E_{g}}}{N}.
\end{equation}
when substituted in the first term in the right hand side of \eqref{eq9} leads to the following second London equation
\begin{equation}
-\upgamma_s {\boldsymbol \nabla} \times {\boldsymbol v}_s = \upgamma_s \boldsymbol{B}_g + \frac{q}{m N}\boldsymbol{H} +
\upgamma_s ({\boldsymbol{E}_g}\times \frac{\boldsymbol{N}}{N})
\end{equation}
in a stationary spacetime in the slicing formalism.
\pagebreak

\end{document}